# NUCLEOSOMES IN GENE REGULATION: THEORETICAL APPROACHES

**V. B. Teif**[1*], **A. V. Shkrabkou**[2], **V. P. Egorova**[2], **V. I. Krot**[3]

[1]*German Cancer Research Center (DKFZ) and Bioquant, 69120 Heidelberg, Germany*
[2]*M. Tank Belarusian State Pedagogical University, Minsk, 220050 Belarus*
[3]*Belarusian State University, Minsk, 220030 Belarus*

This work reviews current theoretical approaches of biophysics and bioinformatics for the description of nucleosome arrangements in chromatin and transcription factor binding to nucleosomal organized DNA. The role of nucleosomes in gene regulation is discussed from molecular-mechanistic and biological point of view. In addition to classical problems of this field, actual questions of epigenetic regulation are mentioned. The authors selected for discussion what seem to be the most interesting concepts and hypotheses. Mathematical approaches are described in a simplified language to attract attention to the most important directions of this field.

***Keywords:*** nucleosome, chromatin, lattice models, competitive binding, transcription factors

*Corresponding author. E-mail: Vladimir.Teif@bioquant.uni-heidelberg.de

## INTRODUCTION

Eukaryotic DNA is packed with the help of nucleosomes, which is known for already 40 years [1]. The importance of the discovery of the nucleosome can be compared with the discovery of the DNA double helix. Indeed, the same year as the historical article of Olins and Olins reporting electron-microscopic visualization of chromatin repeating units appeared in press [1], a similar article by D.F.L. Woodcock was rejected by an anonymous referee with the following comment: "A eukaryotic chromosome made out of self-assembling 70 Å units, which could perhaps be made to crystallize, would necessitate rewriting our textbooks on cytology and genetics! I have never read such a naive paper purporting to be of such fundamental significance" [2]. Today main structural details of the DNA complex with histones in the nucleosome are well established [3]. Nucleosome is composed of the protein octamer containing two pairs of dimers H2A-H2B and H3-H4 wrapped by 146–147 DNA base pairs, which constitutes 1.67 left-handed turns of the double helix. Nucleosomal organization allows packing all human chromosomes (almost two meters of DNA) inside the cell nucleus of a size of just ~10 μm. However, similar packing could be also achieved without nucleosomes. For example, the DNA packing density in unicellular dinoflagellates (which have no nucleosomes) is comparable to that in higher eukaryotes [4]. Apart from DNA compaction, nucleosomes have other, not less important functions. Since about ¾ of genomic DNA is organized by nucleosomes, the majority of regulatory regions are to some extent covered by histones. However, transcription initiation requires a certain arrangement of specific proteins, transcription factors, along these regulatory regions. Nucleosome positions regulate DNA accessibility for transcription factors and RNA polymerases [5–8]. In different situations nucleosomes can impede [9] or facilitate transcription factor binding [10]. Thus nucleosomes are one of the main regulators of transcription. How regulatory proteins bind DNA in the context of nucleosomal organization of the genome? Can nucleosomal DNA be accessible for transcription factor binding? What defines genomic arrangement of nucleosomes? How important is it? How can it be changed? How covalent histone modifications realize epigenetic "memory" of nucleosomes? This mini-review does not pretend to solve these problems, but we will try to outline modern theoretical concepts used to answer these questions. This work complements many recent reviews focused on experimental details of nucleosome-involving processes [11–14, 21–29].

## NUCLEOSOME POSITIONING

A hypothesis that nucleosome positioning *in vivo* is determined by the nucleotide sequence was proposed by Edward Trifonov 30 years ago based on the analysis of a few sets of genomic regions sequenced at that time [15, 16]. It appeared that some dinucleotides are repeated in genomes with a periodicity of ~10 base pairs (bp), which coincides with the period of the double helix. However, a real boom in this area started only recently, after the development of new high-throughput methods allowing determining nucleosome positioning along the whole genome [17–20]. In such experiments DNA between nucleosomes is usually degraded by micrococcal nuclease (MNase), and then remaining nucleosomal DNA is analyzed with the help of microarrays or deep sequencing [21–23]. Recent developments in next generation sequencing created an unprecedented situation when experimental data are accumulating faster than corresponding biophysical models explaining these data [22, 24–29]. It is established that nucleosome positioning in chromatin is determined by three main factors: Firstly, by intrinsic affinity of the histone octamer to the nucleotide sequence [9, 18, 20, 30--33]. Secondly, by the competitive and cooperative binding of transcription factors and other chromatin proteins [27, 34–37]. Finally, ATP-dependent molecular motors, so called chromatin remodelers, can move nucleosomes to new positions or remove them completely [25, 31, 38–41].

## AFFINITY OF HISTONE OCTAMERS TO DNA

At physiological pH and ionic strength the DNA double helix has a persistence length of about 50 nm, which is comparable to the histone octamer size. The rigidity of the double helix is determined to a large extent by the repulsion of negatively charged phosphates. Therefore, significant charge neutralization is required to allow two turns of the DNA in the nucleosome. This neutralization is achieved with the help of positively charged histones. In principle, nucleosomes can be formed at any nucleotide sequence. However, the flexibility of the double helix, which determines the energy of nucleosome formation, depends on the nucleotide sequence. For example, properties of two consecutive nucleotides (dinucleotide) determine the bending probability in a given position along the DNA. Optimally, bending positions would arrange homogeneously along the nucleosomal DNA. Strongest DNA-histone contacts are separated by ~10 bp along the DNA inside the nucleosome [42]. Therefore, optimal nucleosomal sequences are characterized by dinucleotide oscillation with a period of 10 bp [43]. Such features of dinucleotide arrangement have been observed in genomes of most investigated organisms. For example, in the genome of *Saccharomyces cerevisiae*, 14 dinucleotides (all except AC and GT) are repeated with a periodicity ~10.4 bp; in *Drosophila melanogaster* four dinucleotides (AA, TT, CG and GC) are repeated, while in *Homo sapiens* only CG dinucleotides are repeated [44]. This might mean that the role of dinucleotide repeats is probably decreasing with increasing the complexity of the organism. Interestingly, chromatin remodelers also move nucleosomes in steps which are multiples of 10 bp. For example, remodelers NURF and ISW2 can move nucleosomes in step of ~10 bp, while for SWI/SNF the elementary step is around 50 bp [45, 46]. In addition to dinucleotides, several longer sequences such as poly(dA·dT), so called A-tracts, have lower affinity to histone octamer due to their special curved but rigid structure [47]. A-tracts are often found on both sides of a gene in eukaryotic genomes. Besides, a lot of nucleosome-excluding sequences not related to A-tracts have been identified, e.g. $(CCGNN)_n$ [48]. Several sequences with small and large affinity to the histone octamer have been identified *in vitro* [20, 30, 49]. The energy difference of nucleosome formation for genomic sequences is between 0 and 2.4 kcal/mol, while it can reach 4.1 kcal/mol for artificial sequences [50, 51].

Preferred nucleosome arrangements can be predicted from the DNA sequence in the absence of competition of histone octamers with transcription factors and the absence of remodeler action. Several methods of such predictions have been developed in the last years. Biophysical methods are based on the calculation of the flexibility of the double helix composed of different nucleotides, and the corresponding energies of nucleosome formation [35, 42, 52–56]. A second group of approaches is more oriented on bioinformatic methods, where experimentally determined nucleosome positions in the sequenced genomes are used in machine-learning algorithms based on neural networks and/or Markov chains [18, 20, 32, 57–63].

Despite the intuitive similarity of the problem of finding the probability of nucleosome formation to the probability of transcription factor-DNA binding, standard methods cannot be applied to nucleosomes. Typical transcription factors usually cover ~10 bp upon binding to the DNA, which determines $4^{10}$ (more than a million) of possible combinations of four nucleotides A, T, G and C. This number seems large, but it is nevertheless comparable with the number of oligonucleotides which is possible to check in one microarray experiment [64]. Theoretical analysis of such experimental data is usually based on the assumption that the DNA represents a one-dimensional lattice of binding sites (nucleotides, base pairs, dinucleotides, etc), whose action is additive within the protein binding site (the energies of all DNA-protein contacts add up) [64, 65]. Information about the affinity of each protein is stored in the form of position-specific weight matrices (PWM) which allow getting relative binding constants for each nucleotide sequence. For many regulatory proteins such matrices are already determined and systematized in databases such as FlyTF [66], JASPAR [67] and TRANSFAC [68]. However, the nucleosome covers 147 bp, which makes this method not applicable. Indeed, experimental testing of $\sim 4^{147}$ different DNAs of length 147 bp is not possible at a current stage of technology. Therefore, some smart simplifications are needed.

For example, the algorithm of Segal and coauthors [20, 59] sets special weights to a limited number of nucleosome-positioning factors: the dinucleotides mentioned above and 5-nucleotide motifs. The choice of five, not, say, six nucleotides, is purely technical. There are also algorithms where statistical weights are set for tetranucleotide sequences [69]. On the other hand, the algorithm of Trifonov and coauthors [70] considers 10 nucleotides as an elementary motif. This is motivated by the assumption that all positions inside the nucleosome separated by 10 bp are effectively equivalent due to the symmetry of the double helix. In the latter case, mathematical analysis shows that the optimal nucleosomal sequence is $(GGAAATTTCC)_n$, while all other sequences can be considered as deviations from it [29, 33,]. The larger is the deviation, the smaller is the affinity to the histone octamer. Today several online servers exist which allow entering the DNA sequence of interest and getting nucleosome formation probabilities as output [20, 56, 62, 71, 72].

## LATTICE MODELS

Predicting histone octamer affinity to the DNA sequence is only one part of the problem. A second part is to reconstruct the arrangement of transcription factors and nucleosomes in chromatin [73]. The latter problem is usually more difficult. In a general case, each protein is characterized by its molar concentration, site-specific affinities for the DNA and other proteins, and cooperativity parameters. Each protein can occupy one or more DNA lattice units upon binding, with the elementary lattice unit being e.g. a base pair. There are several methods allowing calculating maps of multicomponent cooperative protein-DNA binding. These include the binary variable method, combinatorial method, generating functions method, transfer matrix method and recurrent relations method, as reviewed recently [27, 73]. The recurrent relations method, which belong to the class of dynamic programming algorithms, appeared to be the fastest in the case when nucleosome unwrapping is not considered. Therefore it is currently used in many works on calculation of the arrangement of nucleosomes and regulatory proteins in chromatin [20, 35, 36, 74--78]. Interestingly, the dynamic programming method considering the general case of interactions between proteins separated by several base pairs along the DNA was proposed yet in 1978 by Gurskii and Zasedatelev [79]. The theory of Gurskii and Zasedatelev was applied in a series of publications [27, 80–82] and included in the classical Volkenshtein's textbooks on biophysics. However, this method probably was ahead of its time. The problem is that back in the 1970s, lattice models were used mainly to calculate *in vitro* titration curves, which could be done easily by many methods. On the other hand, today the main challenge is the calculation of protein binding maps for large genomic regions, where the calculation time is becoming the bottleneck. The calculation time of the Zasedatelev-Gurskii approach increases linearly with the DNA length, which makes it probably the fastest algorithm for the given task [20, 35, 36, 74--78].

The nucleosome can be considered in general as one of the many types of protein-DNA complexes in the lattice models. A first lattice model specific for nucleosomes was developed by Kornberg and Stryer [83]. In this model, the number of nucleosomes on the DNA was fixed, but their positions could change. This work predicted that nucleosome positions near the boundaries of the DNA segment oscillate. The periodicity of nucleosome arrangement decays with increasing the distance from the boundary. Such boundary effects are not specific for nucleosomes, and have been also predicted and observed for the general case of protein-DNA binding [84–86]. Recent experiments [53, 87–89] and theoretical investigations [39, 90] confirmed the importance of boundary effects for nucleosome positioning. For example, a DNA region immediately before the transcription start site is usually nucleosome-depleted. This region acts as a barrier which determines oscillatory positioning of neighboring nucleosomes [87, 88]. Even a stronger barrier is formed upon binding of so called insulatory protein CTCF, which can position about 20 nucleosomes [89]. Twenty years ago Kornberg and Stryer proposed that "The binding of a sequence-specific protein to DNA creates a boundary whose effect upon neighboring nucleosomes is of first order near the boundary and decays to lower order with increasing distance from the boundary. The preferential binding of histones to certain sequences is a second order effect, whose influence upon neighboring nucleosomes is then of third order" [83]. As we will see later, the question of the relative importance of these effects is still open today. The model of Kornberg and Stryer using a fixed number of nucleosomes was later extended be Nechipurenko and coauthors [91–93] for the case of a variable number of nucleosomes in analogy with usual DNA-ligand binding. This required the introduction of energetic parameters: the binding constant, effective concentration and contact cooperativity parameter for interaction between DNA-bound histone octamers. Currently similar approaches are used for the analysis of nucleosome positioning genome-wide [18, 20, 27, 39, 94, 95].

## NUCLEOSOME UNWRAPPING

Lattice models described above considered the nucleosome as a single entity. It was assumed that a protein bound to the DNA covers a fixed number of DNA base pairs. That is, the protein is either bound or not, intermediate states being prohibited. For example, if a protein covers $m$ DNA base pairs starting from position $n$, then all base pairs from $n$ to $n + m - 1$ are protected by this protein and cannot be bound to other proteins. However, in reality protein-DNA binding does not happen according to the "all-or-none" model, but goes through several intermediate states. For example, a transcription factor UBF consists of several HMG-domains, which consecutively bind DNA depending on the bend of the double helix introduced by previously bound domains of this protein [96]. Such binding seems to be a rule rather than exception. In particular, gene regulation in chromatin is frequently tuned by changing DNA accessibility for transcription factors through partial unwrapping of the nucleosome [49, 97–106]. Matrix formulations of the lattice models for protein multimer assembly on the DNA allow considering the nucleosome as a particular case of a protein multimer [107], which can partially dissociate when some of histone dimers leave the octamer [27].

Several models have been proposed recently for the description of single-molecule experiments with chromatin fibers, where the DNA chain with nucleosomes is stretched by magnetic tweezers or atomic force microscopy [14, 108–111]. The nucleosomes can be unwrapped in these experiments. However, such descriptions are not applicable directly to nucleosome unwrapping *in vivo*, which can occur spontaneously in the absence of external forces. Therefore, a special one-dimensional lattice model was developed recently. It is based on classical lattice models for DNA-protein binding, but allows describing intermediate states where the nucleosome is partially unwrapped [37]. The idea is that although physically the DNA is wrapped around the histone octamer, mathematically this is equivalent to a situation when a protein complex binds the DNA lattice and covers $m \leq 147$ base pairs. Binding complexes with $m < 147$ bp correspond to partial unwrapping and are quantitatively characterized by statistical weights corresponding to breaking one or more histone-DNA contacts. This model predicted two effects: 1) Partially unwrapped nucleosomal DNA becomes accessible to transcription factor binding, while the central part of the nucleosome is stabilized and becomes less accessible. 2) Partially unwrapped nucleosomes can occupy territories of each other. That is, two neighboring nucleosomes can cover less than $2 \times 147$ bp. Both these effects were indeed observed experimentally [103, 112]. Furthermore, the lattice model taking into account nucleosome unwrapping outperformed the conventional all-or-none binding model as judged by comparison with quantitative *in vitro* AFM measurements of nucleosome positions along a DNA molecule of known sequence [53]. Another type of experiments where DNA accessibility to fluorescently labeled transcription factors was measured *in vitro* as a function of the distance from the DNA entry/exit [102] allowed estimating the unwrapping energy as ~1–2 $k_B T$ per nm DNA, consistent with other theoretical estimates [109]. A transfer matrix model for nucleosome unwrapping is formulated mathematically in such a way that it includes all-or-none binding as a particular case. Therefore, it can be considered as a general update of previous lattice models for nucleosome arrangement on the DNA [37]. A similar model for a simplified system without site-specificity was recently considered by Mirny in the frame of the combinatorial approach [113]. Teif and Rippe also formulated the solution of this problem using the recurrent relations method [73]. It was shown that unlike simpler systems, the computation time of the transfer matrix method and the recurrent relations method become comparable for the case when nucleosome unwrapping it taken into account [73].

## TRANSCRIPTION FACTOR BINDING TO NUCLEOSOMAL DNA

Interaction of transcription factors and nucleosomes is an integral part of any eukaryotic gene regulation process. There exists a specific class of transcription factors which can bind the intact nucleosome without the requirement of its unwrapping [51]. These include, for example, transcription factor FoxA [114]. However, the majority of transcription factors compete for DNA binding with histone octamers. Especially interesting case of such competition is when two or more transcription factor binding sites are situated close to each other. Calculations show that if one transcription factor managed to bind DNA, this stabilizes partially unwrapped nucleosome conformation facilitating binding of the second transcription factor [26, 37, 113]. Since transcription factor binding sites often form clusters in the genome (~10–20 binding sites at a DNA region of several hundred base pairs) [115], such cooperative effects are expected *in vivo* and have been observed *in vitro* [116, 117]. This type of cooperativity is called "collaborative competition" [117]. Calculations performed for short 147 bp DNAs in the presence of histone octamers showed that binding of a single transcription factor is hardly possible for typical energies of small protein-DNA interactions, unless the nucleosome can unwrap. If the nucleosome can unwrap, transcription factor binding probability is increased. If two transcription factor binding sites are within a 60 bp region from each other, the collaborative competition effect takes place, significantly increasing the binding probability for both transcription factors. The role of this effect increases when nucleosome unwrapping is taken into account. Interestingly, increasing the distance between transcription factor binding sites above the half-nucleosome length might lead to the reverse effect. In this case, a partially unwrapped nucleosome can fit between two transcription factor binding sites. If the first transcription factor interacts with the nucleosome cooperatively (stabilizes it), then the second transcription factor is being excluded by the nucleosome. It was proposed that this mechanism might be applicable for transcription factors-antagonists, such as *Drosophila* short-range repressors and activators [137].

Calculations performed for longer genomic regions show that transcription factors whose binding sites form tandems (separated by ~10-20 bp) usually win the competition against the nucleosome [37]. The latter statement does not mean that the probability of transcription factor binding at a given position is higher than the probability of nucleosome formation. It rather states that the probability for a given site to be covered by a nucleosome is significantly smaller than the average nucleosome coverage for this genomic region [37]. The valleys between the peaks obtained by ChIP-seq or ChIP-chip in genome-wide nucleosome positioning experiments are usually interpreted as corresponding to sequences with low nucleosome affinity. However, the analysis mentioned above says that such regions might as well correspond to the clusters of transcription factor binding sites [37]. Experimental data suggest that promoters of active genes are often nucleosome-depleted, while promoters of inactive genes are occupied by nucleosomes, which are being removed by remodelers upon gene activation [118–120]. Many promoters not only contain binding sites for transcription factors and RNA polymerase, but also have a weak affinity for the histone octamer (e.g. contain poly(dA·dT)) repeats [25, 40, 47]. Thus the "chicken or egg" question is still open: It is not clear whether promoter regions are less occupied by nucleosomes due to a lower affinity for the histone octamer, or because of competition with transcription factors.

## NUCLEOSOME REPOSITIONING

The picture outlined above was based on the principles of equilibrium thermodynamics, assuming that nucleosome positions correspond to a thermodynamic equilibrium. In reality, there is no thermodynamic equilibrium in the cellular nucleus. Large viscosity, crowding effects and small copy numbers of many molecules mean that in many cases it is not easy to talk about average concentrations of free molecules and kinetic effects prevail [4]. Even in the absence of a thermodynamic equilibrium, arrangement of transcription factors averaged over a large number of synchronized cells still corresponds to the distribution that one would observe at equilibrium [121]. This is not the case for nucleosomes, which might be trapped in "kinetic traps" in positions far from those favored by the thermodynamic equilibrium [122]. Furthermore, nucleosome positions dynamically change with time. To make it clear, let us look at such an integral parameter as the average DNA length between neighboring nucleosomes [51]. The linker length differs not only in different organisms (~7 bp in *Schizosaccharomyces pombe,* ~18 bp in *S. cerevisiae,* ~28 bp in *Drosophila melanogaster* and *Caenorhabditis elegans*, ~38 bp in *Homo sapiens* [22, 123]), but also in different cell types of the same organism although they share the same DNA sequence (e.g. ~26 bp in human cortical neurons and 60 bp in glial cells) [2]). Furthermore, even cells of the same type are characterized by different linker length at different stages of development [124, 125] and depending on activation of immune response [39, 118].

A lattice binding model was developed to account for kinetic effects in nucleosome positioning. In this model, the nucleosome distribution along the DNA was calculated iteratively using the rules of nucleosome repositioning by chromatin remodelers [39]. Remodeler rules were set using a small number of parameters specific for a given remodeler type: 1) The remodeler step is the distance of nucleosome repositioning in an elementary remodeler reaction. The step is determined by the length of the DNA loop formed by the remodeler. It is currently believed that the mechanism of nucleosome movement without complete dissociation of the histone octamer is based on the translocation of the DNA loop from the beginning to the end of the nucleosome [126]. This step is a multiple of 10 bp for most remodelers. 2) Another parameter determines the probability of nucleosome repositioning from a given position to the left or to the right along the DNA. This parameter depends on the nucleotide sequence and remodeler type. It is assumed that the remodeler integrates two types of signal: its own preferences and intrinsic preferences of the histone octamer to a given DNA site. Calculations involving a large number of iterations showed that the action of a nonspecific remodeler which has no repositioning preferences results in equal spacing between the nucleosomes near the DNA boundary [39]. Such an effect was indeed observed in recent experiments. It appeared that positioning of nucleosomes reconstituted *in vitro* significantly differs from positioning of nucleosomes on the same DNA region *in vivo*. However, when remodelers and ATP were added *in vitro*, the nucleosome oscillatory pattern in the vicinity of transcription start sites was re-established, coinciding with the *in vivo* pattern at these regions [138]. The action of a remodeler, which strengthens existing preferences of the histone octamer to the DNA results in preferential rearrangements of nucleosome-formation probability between existing probability peaks. The action of a remodeler, which has its own strong preferences for nucleosome repositioning results in selective removal of some of the nucleosomes. It was shown that the description of other remodeler types can be reduced to the three ones above. A recent experimental work using knockout of Isw1 remodeler has confirmed these conclusions and detected all these three possible remodeling scenarios *in vivo* [127]. The next step will be to identify remodeler-specific rules for all remodelers. This problem is still waiting for its solution.

## HOW MANY GENOMIC CODES EXIST?

Recently the idea of the "nucleosome code" determining nucleosome positioning in the genome by the DNA sequence has become very popular [15, 20]. However, the number of all possible nucleosome sequences, $4^{147}$, is longer than the length of any genome. Therefore, in practice a nucleosome code based on the analysis of one genome is not always a good prediction for the genome of another organism [40, 128]. There is ongoing discussion with respect to whether the nucleosome code is the main determinant of nucleosome positions *in vivo* [22, 24--26, 29, 40, 129--131]. For example, recent work showed that MNase used in nucleosome positioning experiments for linker degradation actually has its own preferences strongly correlated with the nucleosome code [132]. Therefore it was proposed to use chromatin ultrasonic cleavage instead of the MNase treatment. However, it appeared that DNA degradation by sonication also proceeds non-randomly, depending on the nucleotide sequence [133]. In any case, it is clear that binding of the histone octamer to DNA is, as with other proteins, sequence-specific. The question is to which extent this sequence-specificity is revealed in biological processes. In this regard, it is interesting to compare histones with bacterial proteins H-NS, which have a role analogous to histones in eukaryotes. H-NS binds DNA genome-wide and is also characterized by a weak sequence-specificity [134]. However, the role of such sequence-specificity in prokaryotic gene regulation is not known, while the role of nucleosomes in gene activation/repression has been well established by direct experiments [7]. It seems that the nucleosome code overlaps with many other codes in chromatin. For example, a so called CTCF code determines binding sites for the CTCF insulator proteins, which can organize nucleosomes in their vicinity [135]. In addition, a large wave of recent articles proposes a so called histone code realized with the help of covalent histone modifications [136]. It seems that the histone code makes nucleosomes distinguishable from each other, which can be used as an additional signal for chromatin remodelers with respect to moving or not moving a given nucleosome [27]. Thus, although significant progress has been achieved in this field, many important questions are still open.

We thank Karsten Rippe for fruitful discussions and Yevhen Vainshtein, Fabian Erdel and Jan-Philipp Mallm for comments on the manuscript. The work was supported by the BRFFI (grant B10M-2010), the EpiGenSys grant in the frame of the EraSysBio+ project (BMBF), and the BIOMS Fellowship.